\documentclass[aps,prl,twocolumn,amsmath,amssymb,superscriptaddress]{revtex4-1}
\usepackage{graphicx}
\usepackage{dcolumn} 
\usepackage{graphics}
\usepackage{amsmath,amsfonts,amssymb}

\begin{document}
\title{Device Architecture for Coupling Spin Qubits Via an Intermediate Quantum State}
\author{X. G. Croot$^{*}$\footnote{These authors contributed equally to this work.}}
\author{S. J. Pauka$^{*}$}
\affiliation{ARC Centre of Excellence for Engineered Quantum Systems, School of Physics, The University of Sydney, Sydney, NSW 2006, Australia.} 
\author{J. D. Watson}
\affiliation{Department of Physics and Astronomy, Purdue University, West Lafayette, IN 47907, USA.}
\affiliation{Birck Nanotechnology Center, Purdue University, West Lafayette, IN 47907, USA.}
\author{G. C. Gardner}
\affiliation{Station Q Purdue, Purdue University, West Lafayette, IN 47907, USA.}
\affiliation{Birck Nanotechnology Center, Purdue University, West Lafayette, IN 47907, USA.}
\author{S. Fallahi}
\affiliation{Department of Physics and Astronomy, Purdue University, West Lafayette, IN 47907, USA.}
\affiliation{Birck Nanotechnology Center, Purdue University, West Lafayette, IN 47907, USA.}
\author{M. J. Manfra}
\affiliation{Station Q Purdue, Purdue University, West Lafayette, IN 47907, USA.}
\affiliation{Department of Physics and Astronomy, Purdue University, West Lafayette, IN 47907, USA.}
 \affiliation{School of Materials Engineering and School of Electrical and Computer Engineering, Purdue University, West Lafayette, IN 47907, USA.}
 \affiliation{Birck Nanotechnology Center, Purdue University, West Lafayette, IN 47907, USA.}
\author{D. J. Reilly$^\dagger$}
\affiliation{ARC Centre of Excellence for Engineered Quantum Systems, School of Physics, The University of Sydney, Sydney, NSW 2006, Australia.} 
\affiliation{Microsoft Corporation, Station Q Sydney, The University of Sydney, Sydney, NSW 2006, Australia.} 

\begin{abstract}
We demonstrate a scalable device architecture that facilitates indirect exchange between singlet-triplet spin qubits, mediated by an intermediate quantum state. The device comprises five quantum dots, which can be independently loaded and unloaded via tunnelling to adjacent reservoirs, avoiding charge latch-up common in linear dot arrays. In a step towards realizing two-qubit entanglement based on indirect exchange, the architecture permits precise control over tunnel rates between the singlet-triplet qubits and the intermediate state. We show that by separating qubits by $\sim$ 1 $\mu$m, the residual capacitive coupling between them is reduced to $\sim$ 7 $\mu$eV. 
\end{abstract}
\maketitle
Entangling qubits by conditioning the state of one qubit on the state of another is a central requirement of universal quantum computing \cite{divcrit1, divcrit2}. Ideally, two-qubit interactions should be strong, such that entangling gates are fast with respect to single-qubit coherence times, and controllable, to prevent two-qubit interactions from interfering with single-qubit operations.  Direct exchange coupling between neighbouring spins offers a straightforward means of realising fast two-qubit gates \cite{burkard,petta,silicontwoqubit,controlexchange}, however such approaches are challenging since qubits must be positioned in very close proximity to each other \cite{srinivasa} and operated in a way that avoids leakage from the logical qubit space \cite{wardrop,leakage}. An alternative approach is to use the direct capacitive coupling between spin-dependent charge dipoles \cite{Shulman202,weperen,taylornature}, although, at present, this capacitive interaction is relatively weak in comparison to the decohering charge noise of the qubit environment.

The need to overcome these challenges has created significant interest in alternative approaches to entangling gates with spin qubits. Proposals include the use of floating metallic structures \cite{floatinggate}, ferromagnets \cite{ferromagnet}, cavity-mediated interactions \cite{QED,petersson,dispersivewallraff}, crossed Andreev reflection in superconductors \cite{andreev}, surface acoustic waves \cite{SAWtheory,SAWexp,SAWexp2}  and quantum Hall resonators \cite{quantumhall, dohertyqhe}. With many of these schemes, a major driver is the desire to separate qubits, thereby overcoming gate-crowding and unwanted single-qubit crosstalk whilst maintaining control over two-qubit interactions. 

Here, we demonstrate a device architecture which facilitates indirect exchange coupling between two spatially separated singlet-triplet qubits formed in double quantum dots (DQDs). Coupling is mediated by a multi-electron quantum dot acting as a non-computational, intermediate quantum state (IQS) \cite{Malinowski, jelena1, bluhm, srinivasa} as shown in Fig. 1(a). Overcoming the challenge of loading and unloading electrons in linear arrays \cite{tarucha5dot}, we position accumulation gates over the tunnel barriers to the IQS, allowing transfer of electrons to and from the IQS independently of adjacent qubits, without charge-latching. This architecture also enables qubit interactions to be controlled, either by opening and closing tunnel barriers to the IQS or by modulating its chemical potential.

Mechanisms for coupling spin qubits via an IQS include direct exchange \cite{srinivasa,rkkycharlie}, super-exchange \cite{superexchange, taruchaexchange}, virtual population \cite{virtualexchange,bluhm} and the Ruderman-Kittel-Kasuya-Yosida (RKKY) interaction \cite{rkkycharlie}. 
Regardless of the specific coupling mechanism, a key requirement is the independent loading of qubits and precise control over the tunnel rates between the IQS and adjacent quantum dots. In the present work our focus is coupling two-electron singlet-triplet qubits, although we note that our device can also be configured to couple single spins.

\begin{figure*}
\includegraphics[width=1\textwidth]{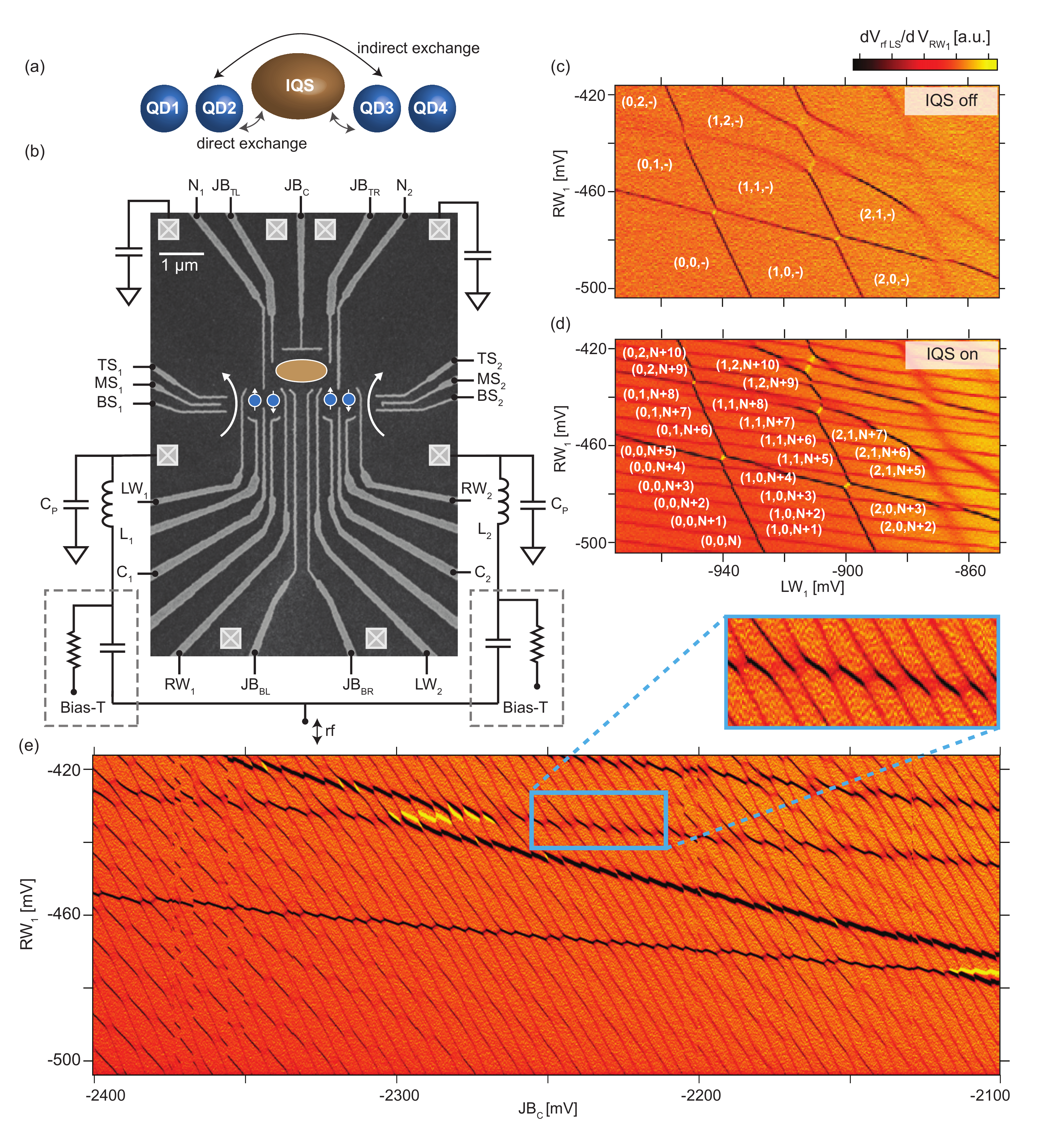}
\caption{ $\bf{(a)}$ Cartoon of our device architecture for coupling singlet-triplet qubits via an intermediate quantum state. $\bf{(b)}$ Electron micrograph and circuit schematic of the double quantum dots (blue circles) coupled by an intermediate quantum state (brown `jellybean' ellipsoid). Crosses indicate ohmic contacts. Inductors ($L_1$ and $L_2$), in resonance with the parasitic capacitance ($C_p$), for tank circuits for impedance matching. $\bf{(c)}$ Charge stability diagram measured with the left sensor as a function of the gates on the left double dot, and without the presence of the dot associated with the IQS, and $\bf{(d)}$ with the IQS configured using gates  $JB_C$, $JB_{BL}$, $JB_{BR}$ and $N_2$. Labels  indicate the charge states (see text). $\bf{(e)}$ The occupation of the intermediate quantum state can be controlled via $JB_{C}$. Transitions associated with the left double dot are seen as the two darker transitions with the shallowest slope. (Inset) Transitions between QD2 and the IQS exhibit a curvature at their triple points, characteristic of level repulsion.}
\label{figure1}
\end{figure*}

The DQDs and IQS are formed in a two dimensional electron gas located 91 nm below the surface of a GaAs/AlGaAs heterostructure (with a density of 1.5 $\times$ $10^{11}$ cm$^{-2}$, and a mobility of 2.4 $\times$ $10^6$ cm$^2$/Vs). Hafnium oxide, deposited using atomic layer deposition, separates TiAu gates from the heterostructure and enables positive voltages to be applied without gate-leakage. Experiments were performed in a dilution refrigerator at a base temperature of 20 mK, with an electron temperature $T_{e} \sim$  90 mK. All data presented was taken in the presence of an applied parallel magnetic field of $B$ = 100 mT. A scanning electron micrograph of the device is shown in Fig.  1(b). To define the DQDs [shaded blue in Fig. 1(b)] we use a well-established gate configuration, known to produce configurable qubits in isolation \cite{hornibrook,croot}. This gate configuration allows each double-dot to be tuned independently prior to coupling via the IQS. 

The IQS is a large, multi-electron quantum dot, configured using both depletion ($N_{1}$, $N_{2}$, $JB_{C}$, $JB_{BL}$, $JB_{BR}$), and accumulation gates ($JB_{TL}$ and $JB_{TR}$).  Positive voltages applied to the accumulation gates control the tunnel barriers between the leads and IQS and ensure that coupling of the $JB_{C}$ gate to the IQS - DQD tunnel barriers can be compensated, such that the barriers remain sufficiently transparent. Readout is performed via rf-reflectometry \cite{readoutreilly}, using an rf-quantum point contact (rf-QPC) to sense the left DQD and an rf-sensing dot (rf-SET) to sense the right DQD. The demodulated reflectometry signal, $V_{rf}$, is proportional to the conductance of the sensors, determined by the charge configuration of the multi-dot system.

We first independently tune both the left and right DQDs into the single electron regime, measuring typical charge stability diagrams, such as what is shown in Fig. 1(c) for the left DQD, where the notation ($n,m,k$) refers to the number of electrons in each triplet-dot system, with $k$ or $n$ indicating the number in the IQS, when referring to the left or right DQD respectively. We next bring up the IQS, and configure it in a regime where there is tunnelling into both the IQS and DQDs. This is straightforward since both charge sensors, positioned at the ends of the array, are sensitive to charge transitions of the IQS as well as their proximal DQD, as seen in Fig. 1(d) for the left sensor. Here, charge transitions of the IQC can be seen overlaying the familiar honeycomb charge stability diagram of the DQD. Note that the presence of the IQS, when configured appropriately with accumulation gates,  hardly shifts the gate voltages needed to define the left DQD. Furthermore, the number of electrons in the IQS can be independently controlled using  gate $JB_{C}$, as shown in Fig. 1(e), which plots the left sensor signal as a function of voltage applied to $JB_{C}$. 

Close inspection of Fig. 1(e) shows several near-horizontal lines that correspond to charge transitions on the left DQD,  whereas the more vertical transitions correspond to the IQS. The IQS occupancy is tunable over a range of at least 50 electrons, and for certain values of $JB_{C}$, the transitions alternate between the signatures of a single and a double dot. In what follows we operate the IQS as a single quantum dot, but note the potential for more complicated interactions when the IQS itself comprises a tunnel-coupled double dot. 
 
\begin{figure}
\includegraphics[width=0.5\textwidth]{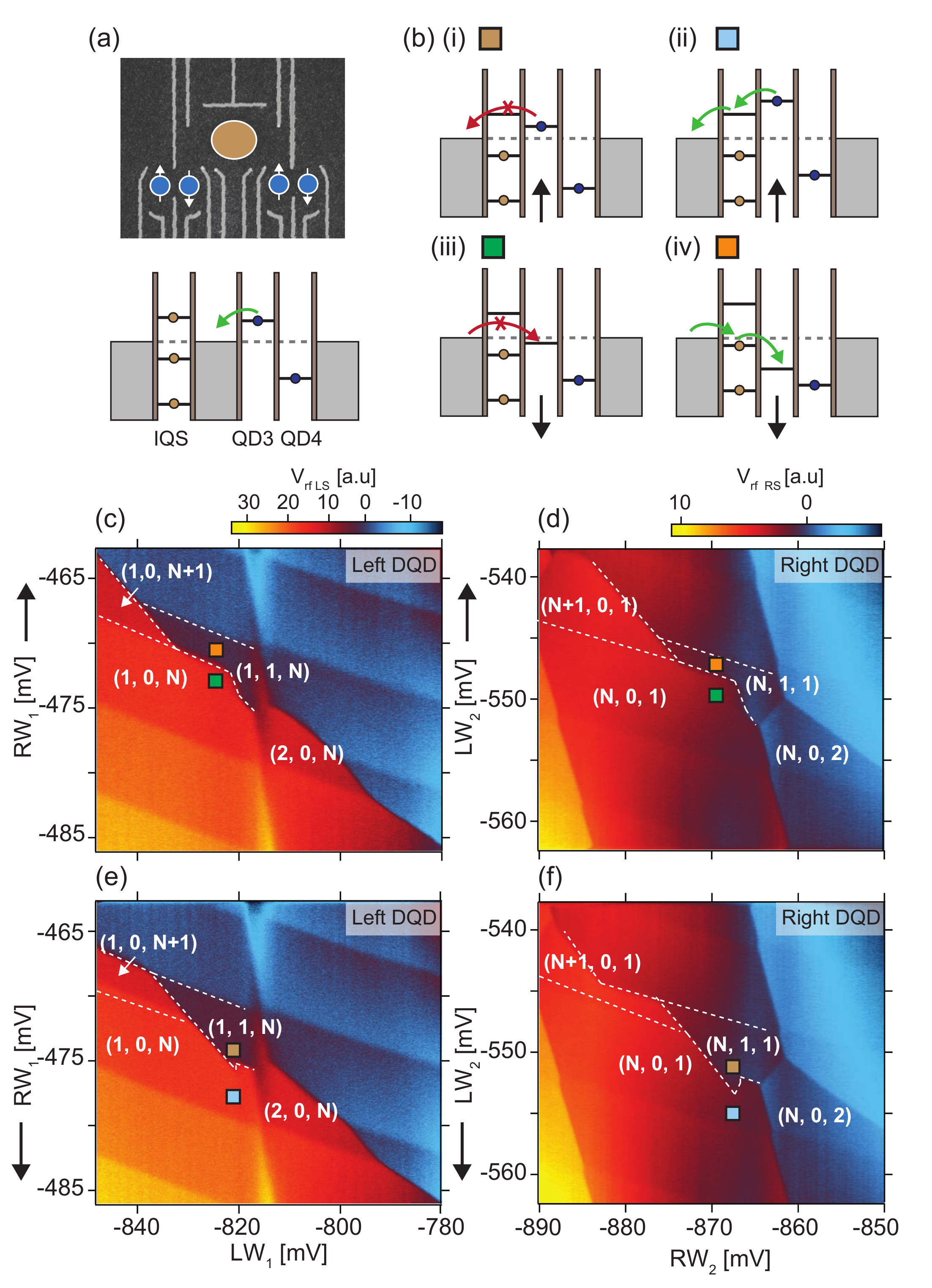}
\caption{\label{}  $\bf{(a)}$ Micrograph of device is displayed for reference. $\bf{(b)}$ Chemical potentials of the multi-dot system when electrons can tunnel directly from the IQS or DQD to a reservoir. (i)-(iv) In the case that the IQS and DQD are tunnel coupled and the barriers are sufficiently opaque, the inner dots (QD2 and QD3) unload via the IQS or outer dot. When the potential of an inner dot is rapidly swept, its charge state depends on whether it is being loaded or unloaded. Unloading ((i) and (ii)), is facilitated by elastic tunnelling through excited or empty states on the IQS, while loading ((iii) and (vi)) can occur once the inner dot potential falls below the IQS potential. $\bf{(c)}$ $\&$ $\bf{(e)}$ Charge stability diagram for the left double dot (sensed with left sensor), as RW${_1}$ is swept from more (less) to less (more) negative. Coloured boxes correspond to the potential configurations in (b). $\bf{(d)}$ $\&$ $\bf{(f)}$ Charge stability diagram for the right double dot (sensed with right sensor), sweeping LW${_2}$ from more (less) to less (more) negative.}
\label{figure2}
\end{figure}

The data in Figs. 1(d) and (e) indicate that the charge transitions in the left DQD undergo level-repulsion with the states of the IQS, although it is not clear from this data whether this interaction is simply capacitive or also involves tunnel coupling (i.e. quantum fluctuations) between the states of the DQD and IQS, which is needed for exchange coupling of the spin states \cite{dassarma}. The picture is made more difficult to interpret since (intra-dot) tunnel transitions between any of the dots and the leads  will also modify the energy levels of the system.  Separating capacitive and tunnel contributions, both inter-dot and intra-dot, is possible by extracting the width of the transitions, as well as by observing how occupancy of the dots depends on the sweep direction of the gates that control the chemical potential. 

We first examine tunnelling between the right DQD and the IQS, in the regime where inelastic tunnelling between the inner dot (QD3) and leads is suppressed. When the potential of QD3 is rapidly increased, it is energetically unfavourable for electrons on QD3 to tunnel to the reservoirs, except via inelastic tunnelling through unoccupied excited states of the IQS, as shown in (i) of Fig. 2(b). Tunnelling out to the lead will therefore not occur for QD3 until the effective triple dot is configured such that the IQS excited state is accessible, as indicated in (ii) in Fig. 2(b). Similarly, depending on whether the energy level of the inner dot is increasing or decreasing, the conditions for loading or unloading electrons via the IQS will differ, as shown in (iii) and (iv) of Fig. 2(b). We look for the signatures of these conditions in the charge stability diagrams for both the left and right DQDs, making use of the corresponding left and right charge sensors. 

The stability diagrams for both the left and right DQDs are acquired using fast charge-sensing by rapidly sweeping the gate that corresponds to the vertical-axis of each data-set from negative to positive, [Fig. 2(c) and (d)], and from positive to negative [Fig. 2(e) and (f)]. The gate indicated on the vertical-axis couples most strongly to the respective inner dot. Comparing Figs. 2(c) and (d) for the left and right DQDs respectively, we see that IQS transitions modify the bare DQD charge stability diagram such that it resembles the diagram expected for a triple-dot system \cite{tripledotstab}. Furthermore, we find that this triple-dot pattern now appears different for opposite directions of the gate sweep. This directional dependence arises when considering the different gate-bias conditions under which the dots will be in a stable occupancy configuration, loading and unloading via tunnelling through states in the IQS [as indicated in Fig. 2(b)]. 
\begin{figure}
\includegraphics[width=0.45\textwidth]{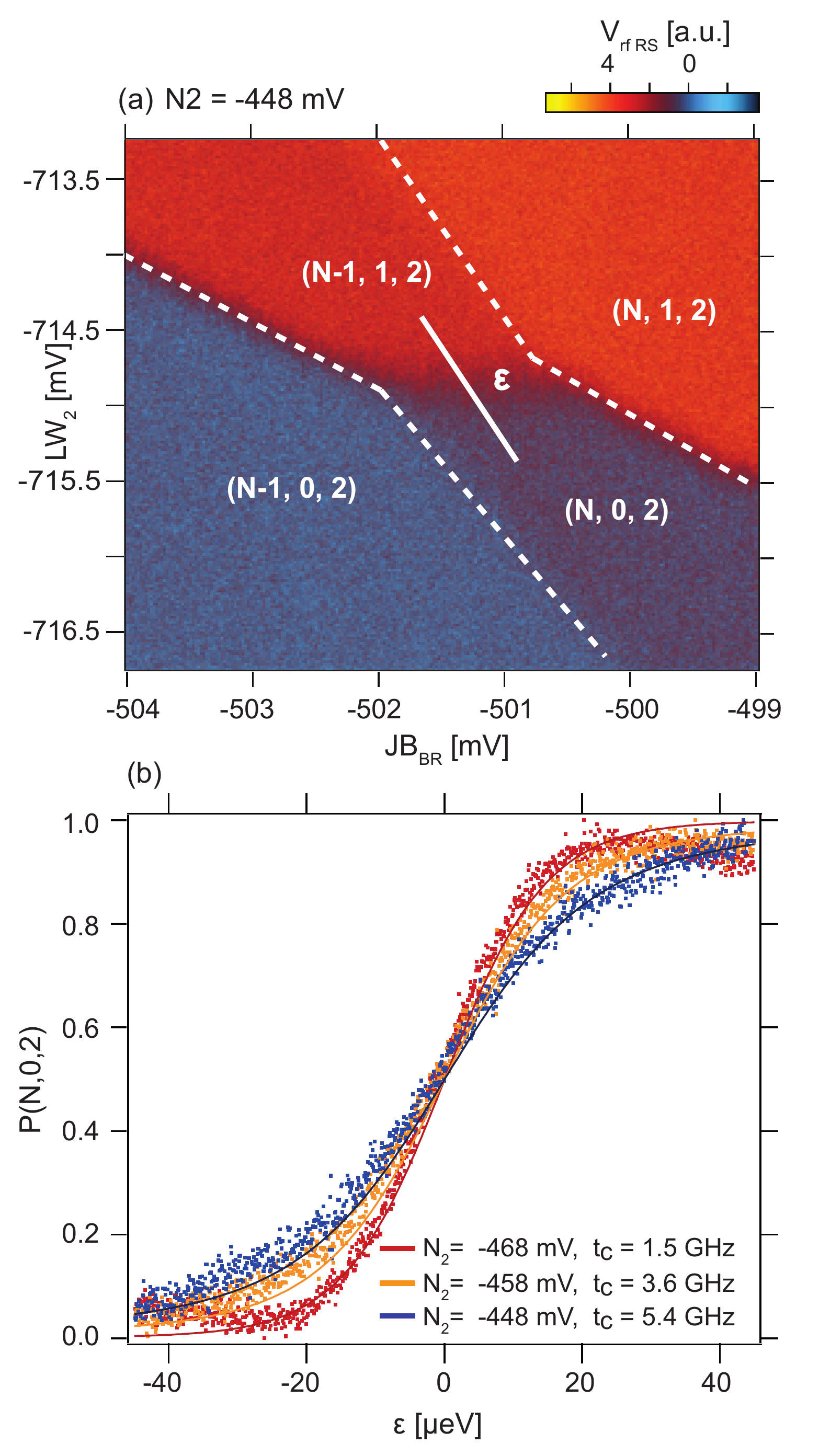}
\caption{\label{} $\bf{(a)}$ Charge stability diagram of the intermediate quantum state (IQS) and right DQD, detected with the right charge sensor. $\bf{(b)}$ Line-cuts taken through the inter-dot transition, along the axis of detuning $\epsilon$ for different voltages applied to gate $N_2$. Tunnel couplings $t_c$ are extracted.}
\label{figure3}
\end{figure}

Having demonstrated that the DQDs and IQS can be configured such that tunnelling occurs between the DQDs and IQS, we now turn to controlling this tunnel rate. The magnitude of the effective exchange interaction between two singlet-triplet qubits separated by an IQS can be tuned by controlling tunnelling rates between the inner dots and IQS \cite{bluhm,superexchange,srinivasa}.  Zooming-up on the charge stability diagram, a transition from the states (N,0,2) to (N-1,1,2) is shown in Fig. 3(a), where the diagonal line indicates the axis of detuning $\epsilon$ between the two states. Plotting the normalized probability $P$ of occupying the (N,0,2) state as a function of detuning, Fig. 3(b) shows how the width of the transition is controllable by altering the bias applied to gate $N_2$. Fitting to these transitions \cite{dicarlo}, we extract the tunnel couplings between the IQS and DQD,  which varies from 5.4 GHz to 1.5 GHz with a 20 mV change in $N_2$ \footnote{Some uncertainty in our estimate of the lever-arm is likely since we have assumed that it is comparable to lever-arm extracted for the right DQD. This is a reasonable assumption given the similar sizes and geometries of the dots.}.

\begin{figure}
\includegraphics[width=0.5\textwidth]{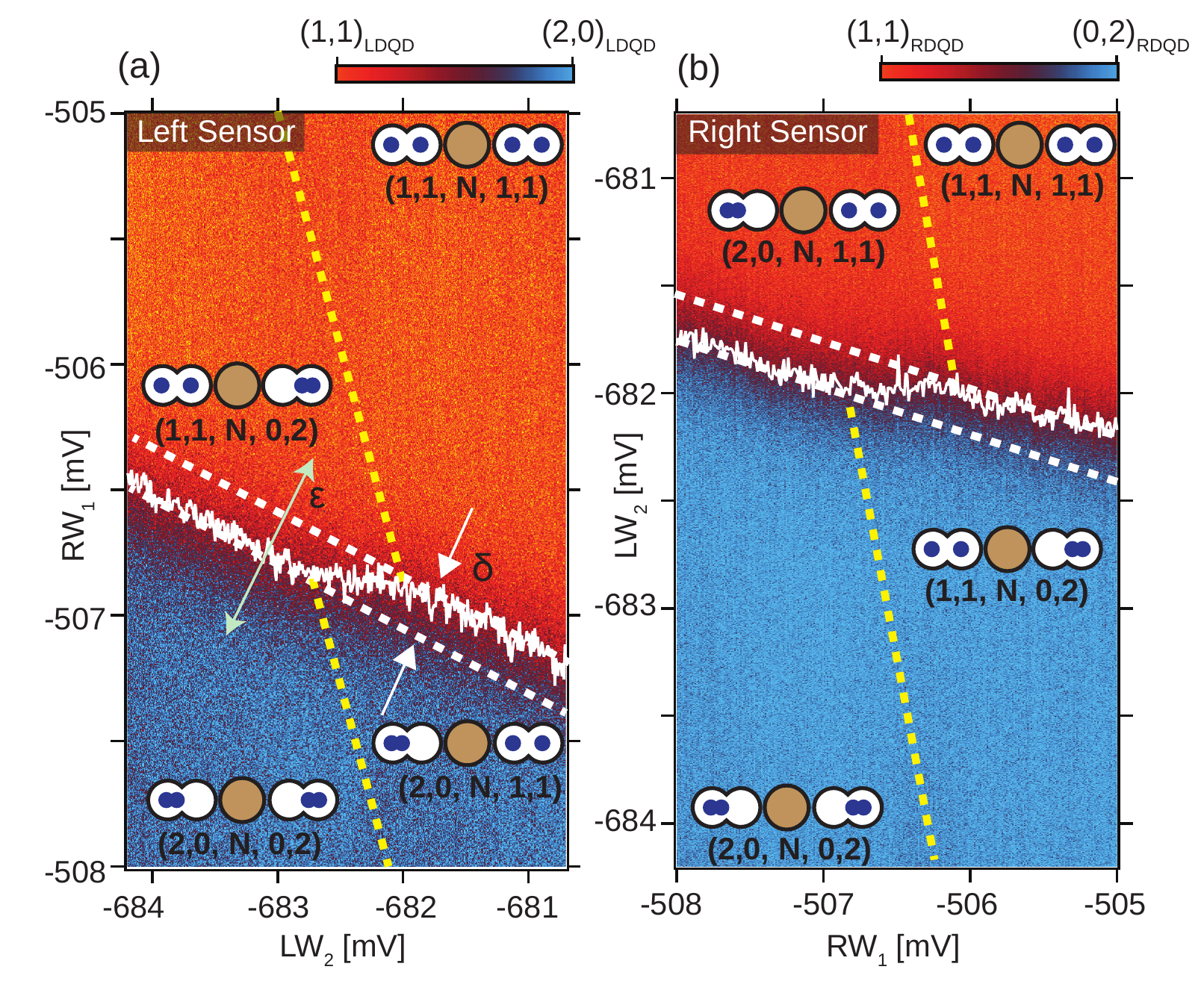}
\caption{\label{} $\bf{(a)}$ Transition of the target DQD, from (1,1) (red) to (0,2) (blue) as the control DQD is configured to switch state from (1,1) to (2,0). DQD configuration detected with the left sensor. The white line corresponds to the position, in gate-space, where $\epsilon$ = 0. Dashed yellow and white lines are guides to the eye: white lines indicate the continuation of $\epsilon$ = 0 in the absence of a second DQD, while yellow dashed lines indicate the approximate $\epsilon$ = 0 coordinates on the opposite double dot.  $\bf{(b)}$  Same as (a), but on the right DQD, measured with the right charge sensor. Using lever arms from both double dots, and appropriate projections, we estimate an electrostatic coupling energy of  6.0 $\mu$eV on the left DQD, and 7.7 $\mu$eV on the right DQD.}
\label{figure4}
\end{figure}

For singlet-triplet qubits coupled via direct exchange, the requirement of tunnel coupling necessitates that quantum dots are in close proximity, where their charge dipoles will also couple via the bare capacitive interaction. In our indirect-exchange architecture, the use of the IQS allows the qubits to be separated by a distance that diminishes their electrostatic coupling such that 2-qubit interactions can be effectively switched-off via gate control of the IQS. To further evaluate our architecture we next measure the residual capacitive coupling when the DQDs are separated by $\sim$ 1 $\mu$m, 10 times the typical distance of qubits engineered with intentionally strong capacitive-coupling \cite{weperen,Shulman202}. Tuning the IQS into the Coulomb blockade regime, we configure both DQDs as singlet-triplet qubits operating on the two-electron system that spans the (0,2)-(1,1) charge states. To measure the capacitive coupling, Figs. 4(a) and 4(b) show the response of each sensor, with the colour scale normalised to configurations of the two-electron charge states of each DQD.

The transition, from orange to blue in Fig. 4(a), corresponds to the left target DQD switching its charge state from (1,1) to (2,0) with detuning $\epsilon$, as measured by the left sensor. Figure 4(b) shows the equivalent transition for the right DQD, from (1,1) to (0,2), now measured with the right sensor. To determine the capacitive coupling between the two dipoles, we looks for a shift in the position of this transition $\delta$ on the left target DQD, as the right control DQD switches between its two charge states. Fitting the position of the transitions with gate voltage (the gate values for which $\epsilon$ = 0) is shown in white in the stability diagrams of Fig. 4(a) and (b) \cite{dicarlo}. Finally, the shift in position of the target transition $\delta$ can be converted to an effective electrostatic energy using the lever-arms separately extracted from bias-spectroscopy measurements of the DQDs. Using this approach we measure a differential cross-capacitive interaction of 6.0 $\mu$eV  when the left DQD is configured as the target and 7.7 $\mu$eV when the target is the right DQD. These energies can be compared to measurements made in ref. \cite{weperen}, where a 100 nm DQD separation yields an interaction energy of $\sim$ 25 $\mu$eV.  We presume that in our device the presence of the IQS, populated with some tens of electrons, accounts for the enhanced capacitive coupling over what may be expected from considering the linear scaling of the bare device geometry. 

In summary, we have presented a device architecture that enables independent loading and unloading of electrons across five quantum dots using both depletion and accumulation gates to control tunnel barriers. Fast charge sensors, positioned at the ends of the device structure, are shown to be sufficiently sensitive to allow tuning of both DQDs and the quantum dot that is host to the intermediate quantum state. The platform alleviates the burden of spatial-crowding suffered by qubits that are coupled via direct-exchange, and opens a means of scaling spin qubits beyond linear arrays. 

$\dagger$ Corresponding author, email: david.reilly@sydney.edu.au \newline

$*$ These authors contributed equally to this work. \newline

This research was supported by Microsoft Station-Q, the US Army Research Office grant W911NF-12-1-0354, the Australian Research Council Centre of Excellence Scheme (Grant No. EQuS CE110001013). We thank A.C. Mahoney for technical assistance and J.M. Hornibrook for the development of the readout multiplexing chips.

 
%\bibliography{fivedot}

\end{document}